\documentstyle[prl,aps]{revtex}
\draft
\input epsf
\input{epsf.sty}
\begin{document}

\twocolumn[\hsize\textwidth\columnwidth\hsize\csname
@twocolumnfalse\endcsname

\title{
 Strong quantum fluctuation of vortices in the new superconductor $MgB_2$}

\author{
H. H. Wen\cite{responce}, S. L. Li, Z. W. Zhao, Y. M. Ni, Z. A. Ren, G. C. Che, H. P. Yang, Z. Y. Liu and Z. X. Zhao}

\address{
National Laboratory for Superconductivity,
Institute of Physics and Center for Condensed Matter Physics,
Chinese Academy of Sciences, P.O. Box 603, Beijing 100080, China
}

\maketitle

\begin{abstract}
By using transport and magnetic measurement, the upper critical field $H_{c2}(T)$ and the irreversibility line $H_{irr}(T)$ has been determined. A big separation between $H_{c2}(0)$ and $H_{irr}(0)$ has been found showing the existence of a quantum vortex liquid state induced by quantum fluctuation of vortices in the new superconductor $MgB_2$. Further investigation on the magnetic relaxation shows that both the quantum tunneling and the thermally activated flux creep weakly depends on temperature. But when the melting field $H_{irr}$ is approached, a drastic rising of the relaxation rate is observed. This may imply that the melting of the vortex matter at a finite temperature is also induced by the quantum fluctuation of vortices.    
\end{abstract}


]
The recently discovered new superconductor $MgB_2$ generates enormous interests in the field of superconductivity\cite{akimitsu}. Several thermodynamic parameters have already been derived, such
as the upper critical field $H_{c2}(0)$ = 13 - 20.4 T\cite{bud1,canfield,takano,finnemore}, the Ginzburg-Landau parameter $\kappa \approx$ 26\cite{finnemore}, and the bulk critical superconducting current density $j_c(0)\approx 2 \times 10^4 A/cm^2$ at 20 K and 1 T\cite{takano,bugoslavsky}. One big issue concerns however in which region on the field-temperature ( H-T ) phase diagram it can carry large critical current density ( $j_c$ ) and thus can be used in the future for industry. This $j_c$ is controlled by the mobility of the magnetic vortices, and vanishes at the melting line between the vortex solid and liquid. This melting can be induced by strong fluctuation of the vortex position by either thermal effect or quantum effect. At T = 0 K only the quantum fluctuation is left. In this Letter we show the evidence for a quantum fluctuation effect at zero K. Further analyses on the magnetic relaxation at finite temperatures reveal that the melting of vortex solid at a finite temperature is also governed by this effect posing a strong limit to the application of this new superconductor in a high field.

Samples investigated here were fabricated by high pressure synthesis ( P = 6 GPa at 950$^\circ$ C for 0.5 hour ) which was described very clearly in a recent publication\cite{ren}. High pressure synthesis is a good technique for producing the $MgB_2$ superconductor since it can make the sample more dense and also prevent the oxygenation of Mg element during the solid reaction.   Our samples are very dense and look like metals with shiny surfaces after polishing. X-ray diffraction ( XRD ) analysis shows that they are nearly in a single phase with the second phase ( probably $MgO$ or $MgB_4$ ) less than 1 wt.\%. Resistive transition was measured by the standard four-probe technique and the magnetic measurements were carried out by a superconducting quantum interference device ( SQUID, Quantum Design MPMS 5.5 T ) and a vibrating sample magnetometer ( VSM 8T, Oxford 3001 ). To precisely calculate the critical current density $j_c$ the sample has been cut with a diamond saw into a rectangular shape with sizes of 4 mm ( length ) $\times$ 3 mm ( width ) $\times$ 0.5 mm ( thickness ).  Fig.1 shows the diamagnetic transition of one of the samples measured in the field-cooling ( FC ) and zero-field-cooling ( ZFC ) process. In the FC process, the temperature was lowered from above $T_c$ to a desired temperature below $T_c$ with a magnetic field, and the data is collected in the warming up process with field. Its signal generally describes the surface shielding superconducting current and the internal frozen magnetic flux profile. In the ZFC process, the temperature was lowered from above $T_c$ to a desired temperature below $T_c$ at a zero field and the data is collected in the warming up process with field. Its signal generally describes the internal magnetic flux profile which is ultimately related to the flux motion. It can be seen from Fig.1 that the transition is very sharp with a perfect diamagnetic signal. The inset shows the resistive transition with $T_c^{onset}$ = 40.3 K and the transition width only 0.4 K. In Fig.2a 
\begin{figure}[h]
	\vspace{10pt}
    \centerline{\epsfxsize 8cm \epsfbox{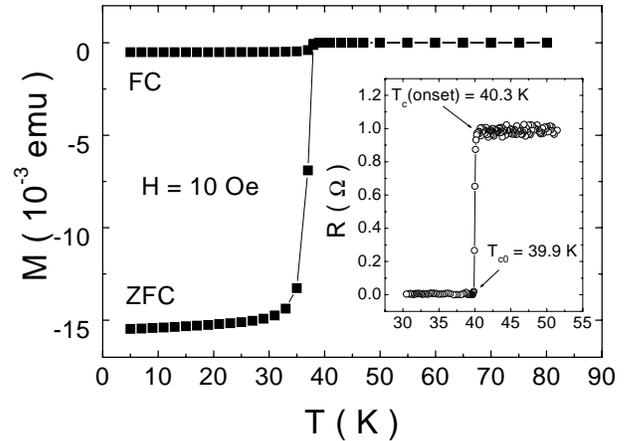}}
    \vspace{10pt}
\caption{Temperature dependence of the superconducting diamagnetic moment measured in the ZFC and FC processes at a field of 10 Oe. A perfect diamagnetic signal can be observed here. The inset shows the resistive transition with $T_c^{onset}$ = 40.3 K and $T_{c0}$ = 39.9 K, the transition width is only about 0.4 K.
}
\label{fig:Fig1}
\end{figure}
\noindent  we show the magnetization hysteresis loops ( MHL ) measured at different temperatures. The symmetric MHLs observed for temperatures up to 38 K indicate the dominance of the bulk current instead of the surface shielding current. The MHL is closed at 8 T for all measured temperatures. Another striking feature is that the MHLs measured at low temperatures ( e.g., 1.6 K, 2.6 K and 5 K ) are too close to be distinguishable. From these MHLs one can calculate $j_c$ via $j_c = 20 \Delta M/Va(1-a/3b)$ based on the Bean critical state model, where $\Delta M$ is the width of the MHL, V, a and b are the volume, length and width of the sample, respectively. The result of $j_c$ is shown in Fig.2b. It is clear that the bulk critical current $j_c$ of 
\begin{figure}[h]
	\vspace{10pt}
    \centerline{\epsfxsize 8cm \epsfbox{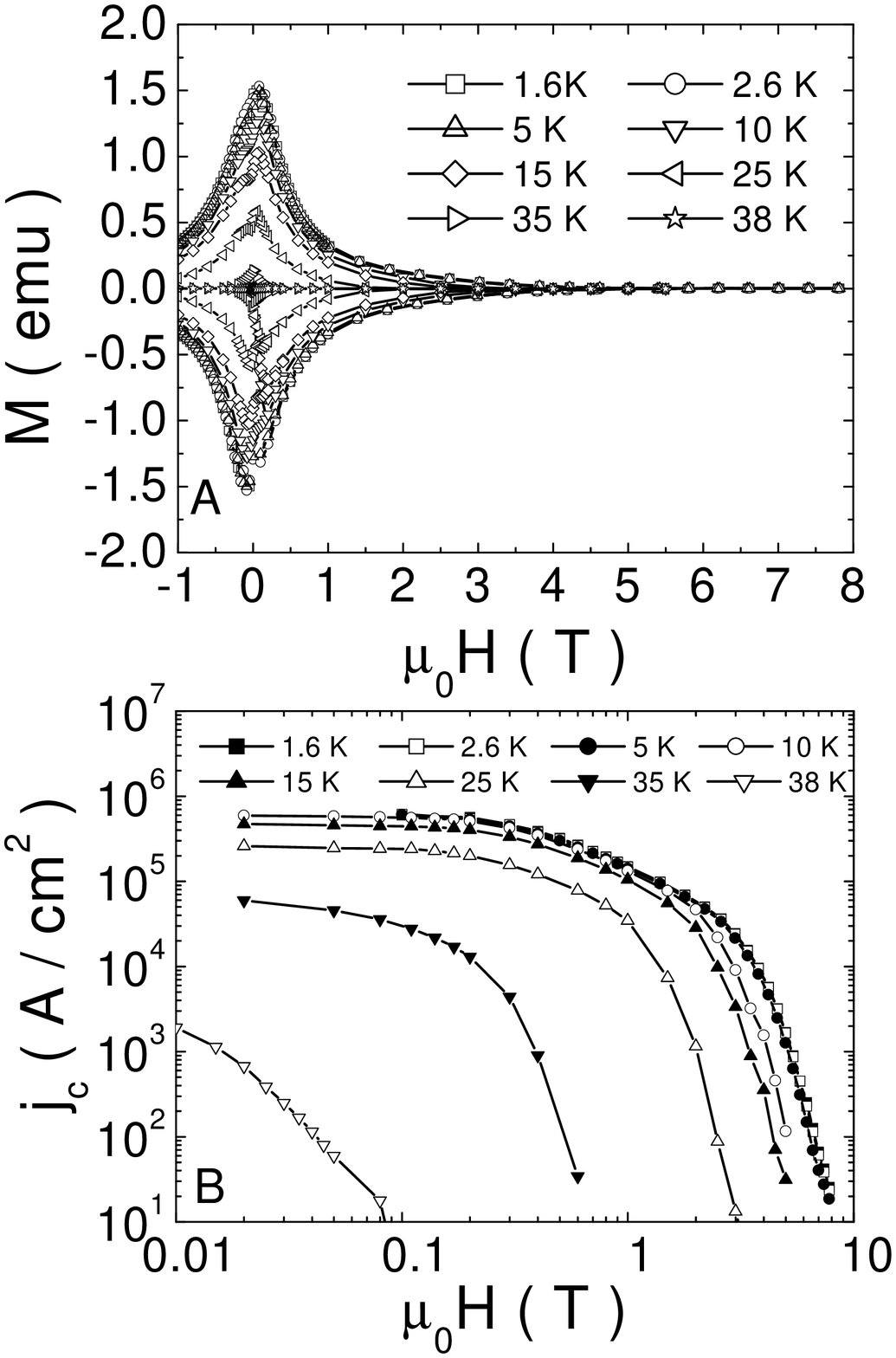}}
    \vspace{10pt}
\caption{(a)Magnetization hysteresis loops measured at 1.6, 2.6, 5, 10, 15, 25, 35, 38 K ( from outer to inner ). All curves here show a symmetric behavior indicating the importance of bulk current instead of surface shielding current. The MHL is closed at about 8 T for all measured temperatures. Another striking feature is that the MHLs measured at low temperatures ( e.g., 1.6 K, 2.6 K and 5 K ) are too close to be distinguishable. (b), Critical current density $j_c$ calculated based on the Bean critical state model. The $j_c(H)$ curves measured at low temperatures are very close to each other showing a rather stable value of $H_{irr}$ when T approaches zero K.
}
\label{fig:Fig2}
\end{figure}
\noindent our sample is rather high. For example, at T = 15 K and $\mu_0H$ = 1 T, we have $j_c = 1.05 \times 10^5 A / cm^2$, which is among the highest values that have been reported\cite{takano,bugoslavsky}.stable value of $H_{irr}$ when T approaches zero K. As mentioned above, the $j_c$ ( $\propto \Delta M$ ) generally reflects the irreversible flux motion, therefore on the melting line between a vortex solid and a liquid, both $j_c$ and $\Delta M$ will vanish. In this way we can determine the melting line or the so-called irreversibility line $H_{irr}(T)$. We have determined the irreversibility line $ H_{irr}(T)$ by taking a criterion of 30 $A / cm^2$. Again the $j_c(H)$ curves measured at low temperatures are very close to each other showing a rather  Another way to determine the $H_{irr}(T)$ and the upper critical field $H_{c2}(T)$ is to measure the temperature dependence of magnetization in the field cooling ( FC ) and zero-field-cooling ( ZFC ) process as shown in Fig.3. A typical example for how to determine $H_{irr}(T)$ and $H_{c2}(T)$ is shown in the inset of Fig.3. The $H_{c2}(T)$ is determined from the point at which the magnetization start to deviate from the normal state linear background, while the $H_{irr}(T)$ line is determined by taking the deviating point between ZFC-FC M(T) curves with a criterion of $\Delta M_{ZFC-FC} = 10^{-4}$ emu. Both $H_{c2}(T)$ and $H_{irr}(T)$ lines are shown in Fig.4a. Also shown in Fig.4a is the $H_{c2}(T)$ determined by Takano et al.\cite{takano} on high pressure synthesized samples. The $H_{irr}(T)$ data determined from both methods mentioned above coincide very well. The $H_{irr}(T)$ determined here is almost identical to that 

\begin{figure}[h]
	\vspace{10pt}
    \centerline{\epsfxsize 8cm \epsfbox{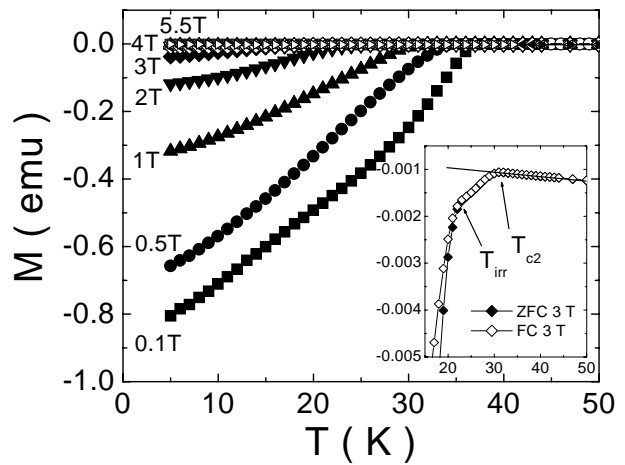}}
    \vspace{10pt}
\caption{Temperature dependence of the magnetization measured in the ZFC and FC process at fields of 0.1 T, 0.5 T, 1 T, 2 T, 3 T, 4 T and 5.5 T. All filled symbols are corresponding to the magnetization measured in the ZFC process. In the main frame all the FC M(T) curves are crowded in a small region. As shown in the inset, the $H_{c2}(T)$ is determined from the point at which the magnetization starts to deviate from the normal state linear background, while the $H_{irr}(T)$ line is determined by taking the deviating point between ZFC-FC M(T) curves with a criterion of $\Delta M_{ZFC-FC}$ = $10^{-4} emu$.}
\label{fig:Fig3}
\end{figure}

\noindent reported by Bugoslavsky et al.\cite{bugoslavsky} and also close to that by Finnemore et al.\cite{finnemore} showing the generality. A striking result is that the $H_{irr}(T)$ extrapolates to about 8 T at zero K, while the $H_{c2}(T)$ extrapolates to a much higher value\cite{hc2} at zero K. There is a large separation between the two fields $H_{c2}(0)$ and $H_{irr}(0)$. If following the hypothesis of the vortex liquid above $H_{irr}$, we would conclude that there is a large region for the existence of a quantum vortex liquid at zero K. This can be attributed to a quantum fluctuation effect in $MgB_2$. Although the lowest temperature in our present experiment is 1.6 K, however, from the experimental data one cannot find any tendency for $H_{irr}(T)$ to turn upward rapidly to meet the $H_{c2}(0)$ at zero K. One may argue that the $H_{irr}(T)$ probably can be increased to higher values by introducing more pinning centers into this sample. This is basically correct, but our samples are dense and almost single phased with very high $j_c$ together with very low magnetic relaxation rate ( shown in Fig.4b ), all these imply strong pinning. 

\begin{figure}[h]
	\vspace{10pt}
    \centerline{\epsfxsize 8cm \epsfbox{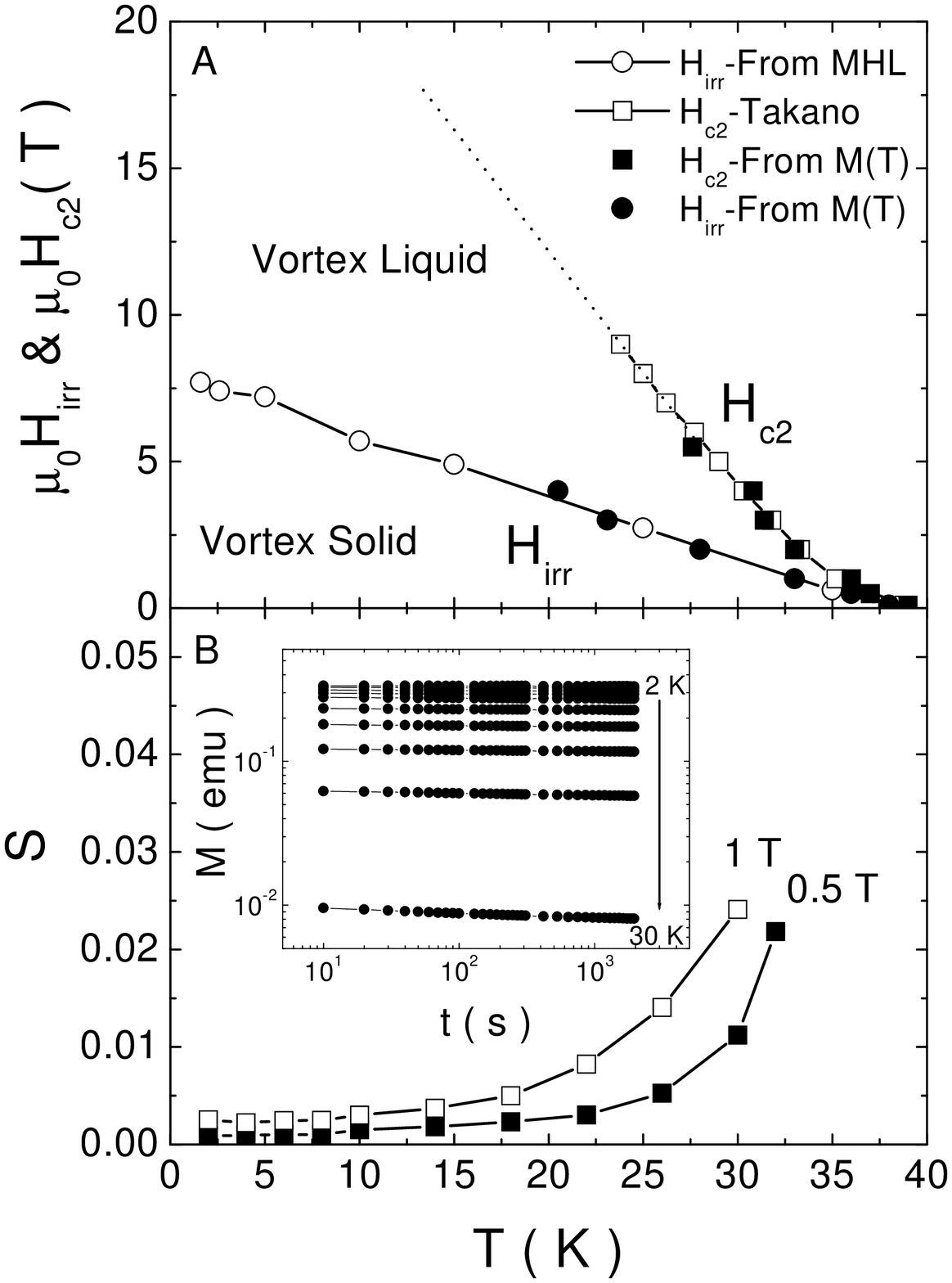}}
    \vspace{10pt}
\caption{(a) H-T phase diagram for the new superconductor $MgB_2$. The dotted line is a guide to the eyes by following the tendency of the measured data of $H_{c2}$. (b) The magnetic relaxation rate determined by $S = - dlnM/dlnt$ at fields of 0.5 and 1 T. The inset shows the original data of magnetization versus time, a good linear relation has been found between lnM and lnt. It is clear that the relaxation rate is very low not only at T = 0 K but also at finite temperatures indicating a strong vortex pinning.}
\label{fig:Fig4}
\end{figure}
\noindent In addition, as mentioned above, the $H_{irr}(T)$ found in present work is very close ( or identical ) to those by other authors\cite{bugoslavsky,finnemore}, therefore the $H_{irr}(T)$ curve here shows a more intrinsic feature. Recently a small openness of MHL is reported\cite{larbalestier} to remain to a high field leading to the uncertainty in determining the $H_{irr}(T)$ curve. However, it is important to note that this feature may only be a secondary effect induced either by some very tiny regions with vortex slush or by the surface barrier of the tiny grains. The major part of the vortex system melt at $H_{irr}(T)$. Above the $H_{irr}(T)$ one should be able to measure the finite linear resistivity. Very recently, we have carefully measured the $H_{irr}(T)$ for bulk samples prepared by high pressure  ( $j_c^{bulk}( 2K, 0.2T)= 1\times10^6 A/cm^2$ ) and ambient pressure ( $j_c^{bulk}( 2K, 0.2T)= 3\times10^6 A/cm^2$ ) synthesis, all deliver almost the same result: $\mu_0H_{irr}(2K)\approx7.8\pm0.3T$. This may imply a strongly linked current flow between grains\cite{larbalestier} and the $H_{irr}(T)$ measured on different samples are very close to each other. 

Theoretically, quantum melting of the vortex solid has been proposed by some authors\cite{blatter,kramer,ikeda} and preliminarily verified by experiments.\cite{sasaki,okuma}. Solid evidence is, however, still lacking mainly because either the values of $H_{irr}(0)$ and $H_{0}(T)$ are too high to be accessible, such as in the classical Chevrel phase PbMoS system\cite{rossel}, or the separation between them is too small \cite{sasaki,okuma} leading to a difficulty in drawing any unambiguous conclusions. Here we try to have an rough consideration on the quantum melting field $H_m$ proposed by Blatter et al.\cite{blatter} for 2D system, 
 
\begin{equation}
H_m(0)/H_{c2}(0) = 1-1.2exp(-\pi^3C_L^2R_Q/4R_{2D}) 
\end{equation}

where $C_L$ is the Lindermann number, $R_Q = \hbar/ e^2 \approx 4.1 k\Omega $, $R_{2D}$ is the sheet resistance. Since the new $MgB_2$ sample has a much higher charge density and thus a much lower sheet resistivity, according to above relation, $H_m$ should be more close to $H_{c2}(0)$ comparing to $high-T_c$ superconductors ( HTS ). This is in contrast to the experimental observations which may be explained as that the $MgB_2$ is not a quasi-2D system. Another approach was proposed by Rozhkov and Stroud\cite{rozhkov}, 

\begin{equation}
H_m(0)/H_{c2}(0) = B_0/(B_0+H_{c2}(0)) 
\end{equation}

with $B_0=\beta m_pC^2s\Phi_0/4\pi\lambda (0)^2q^2$, where s is the spacing between layers, $m_p$ is the pair mass, q the pair charge ( $approx 2e$ ), C the light velocity, $\lambda(0)$ the penetration depth at zero K, $beta \approx$ 0.1. If comparing again the present new superconductor $MgB_2$ with HTS, $\lambda(0)$, q and $\beta$ are more or less in the same scale, the difference comes from $m_p$ and s. Therefore a preliminary conclusion would be that in $MgB_2$ either the pair mass $m_p$ or the layer spacing s is much smaller than that of HTS.

The strong quantum fluctuation here does not naturally imply a strong quantum creep rate as well as a strong thermally activated flux creep in $MgB_2$. The magnetic relaxation rate has been determined via $S = - dlnM/dlnt$ based on the data of magnetization vs. time measured at 0.5 T and 1 T. As shown in Fig.4(b), the quantum creep rate ( if any ) is only about 0.2\% being an order of magnitude lower than that of HTS. This is probably caused\cite{blatter} by the much lower normal state resistivity $\rho_n$ and the much longer coherence length $\xi$ in $MgB_2$. In addition the relaxation rate at a finite temperature is also very low. For example, at 30 K and 1 T, S is only 2.4\% being one order of magnitude lower than that of HTS at the same reduced temperature t = 30 K / 40 K = 0.75. Actually the melting temperature corresponding to 1 T is 35 K, which indicates a sharp transition at the melting point since S should suddenly rise from 2.4\% at 30 K to 100\% at 35 K. The extremely slow relaxation rate at a finite temperature is probably induced by a strong pinning barrier relative to the thermal energy, i.e., $k_BT << U_p$. Therefore for the new superconductor $MgB_2$ we strongly suggest that the melting between a vortex solid and a liquid is due to strong quantum fluctuation instead of the thermal fluctuation. It is also this quantum fluctuation that kills the critical current density and strongly limits the high power application of the new superconductor $MgB_2$ in high field region although it has a strong pinning.

In conclusion, a large separation between the upper critical field $H_{c2}(0)$ and the irreversible field $H_{irr}(0)$ has been observed. This has been explained as due to strong quantum fluctuation effect at zero K. Furthermore, the magnetic relaxation rate is found to be very low at finite temperatures until the irreversibility line $H_{irr}(T)$ is approached showing a possibility that the vortex melting at a finite temperature is also induced by the quantum fluctuation effect. The reason for the strong quantum fluctuation is unknown yet which warrants certainly further investigations.  

\acknowledgements
This work is supported by the National Science Foundation of China (NSFC 19825111) and the Ministry of Science and Technology of China ( project: NKBRSF-G1999064602 ). HHW gratefully acknowledges Dr. A. F. Th. Hoekstra for fruitful discussions, and continuing financial support from the Alexander von Humboldt foundation, Germany.

\end{document}